\documentstyle[11pt,aaspp]{article}
\begin{document}

\title {New Stars on the Block}
\author{Ray Jayawardhana}
\affil{Harvard-Smithsonian Center for Astrophysics, 60 Garden St., Cambridge, MA 02138\\ Electronic mail: rayjay@cfa.harvard.edu}

\begin{abstract}
Groups of young stars located in the solar neighborhood have
recently received a lot of attention. These stars, which are $\lesssim$
10 million years old, can provide insights into the birth 
of stars and planetary systems. For example, the TW Hydrae
Association is just at the age when planets are believed to form,
while MBM12 represents an earlier stage. In this Perspective,
I briefly discuss recent research in this field. 
\end{abstract}

Much glory in astronomy comes from probing the farthest
reaches of the universe for the most exotic beasts. But poking
around closer to home for more prosaic objects has its own
rewards--such as finding small groups of relatively isolated
young stars destined for the limelight. Over the past few years,
one such group, known as the TW Hydrae Association, has
attracted quite a following, as evidenced by the high attendance of
a special session during the recent American Astronomical
Society meeting.

The reason for this widespread interest is that the TW Hydrae group, and
others like it in the solar neighborhood, may
tell us a lot about the birth of stars and planetary systems. The TW
Hydrae Association, in particular, appears to be at
the age at which planet formation is believed to occur. It may even be
possible to take a picture of a newborn planet
around one of these stars.

The star TW Hydrae first caught astronomers' attention back in 1978 (1).
George Herbig noted that it had the earmarks
of a young low-mass star, or so-called T Tauri star, including variability
in brightness, a strong H$\alpha$ emission line in its
spectrum, and a high abundance of lithium (an element that is easily fused
in nuclear reactions and thus does not
survive in older stars). Most T Tauri stars are found in clouds of gas and
dust, the presumed sites of their birth, such
as the Orion nebula. Curiously, TW Hydrae is not.

Subsequent work added further evidence for TW Hydrae's youth and
indications for the presence of a circumstellar
disk and revealed four other stars in the same region of the sky with
similar characteristics (2, 3). Three years ago,
Kastner et al. (4) suggested on the basis of strong x-ray emission from
all five systems that the group forms a
physical association at a distance of roughly 150 light-years. Since then,
at least seven more stars have been identified
as candidate members of the TW Hydrae Association, on the basis of the
same signatures of youth and the same
motion across the sky as the original five members (5).

The group consists mostly of low-mass stars, typically a few tenths of the
mass of the sun, and includes several binary
systems as well as one remarkable quadruple system, HD 98800, in which two
pairs of stars appear to orbit a common
center of gravity. There is only one higher mass star, HR 4796A, which is
twice as massive as the sun and about 20
times as luminous. The TW Hydrae stars are estimated to be roughly 10
million years (My) old (4, 6), older than most
T Tauri stars in star-forming regions, which are usually only about 1 My
old.

The origin of the TW Hydrae Association remains a bit of a mystery. There
is no obvious parent cloud (1, 2), and the
stars are dispersed across some 20 on the sky and 60 light-years in radial
distance, making it difficult to determine
their birthplace (7). Were they born in a low-mass cloud that has since
dispersed? Or could these stars be escapees
from known star-forming regions (8)? The slow velocities of TW Hydrae
stars through space favor in situ formation,
suggesting that clouds may disperse more quickly than previously thought
(9).

Being the nearest group of young stars (and three times closer than the
nearest previously known star-forming region),
the TW Hydrae Association offers a unique opportunity to study the
evolution of circumstellar disks and planet
formation. Furthermore, its estimated age of 10 million years provides a
strong constraint on disk evolution time scales
and fills a substantial gap in the age sequence between previously known
1-My-old T Tauri stars and 50-My-old
nearby open clusters. It has been suggested that circumstellar disks
evolve from dense, actively accreting structures to
sparse, passive remnants within about 10 My (10). During this transition,
grains may assemble into planetesimals, or
the disk may be cleared by planets. The circumstellar disks of the TW
Hydrae stars exhibit a wide variety, from
classical T Tauri accreting disks, to planetary debris systems, to systems
without measureable disk emission at
near-infrared wavelengths implying cleared-out inner disks (11). A
spectacular debris disk with a central cavity has
been directly imaged around HR 4796A (12, 13). The diverse disk properties
suggest that the TW Hydrae stars are at
an age when disks are rapidly evolving through coagulation of dust and
dissipation of gas.

If planets have indeed formed around these stars, it may be possible to
detect them with large ground-based telescopes.
Adaptive optics, a technique that corrects for the blurring of the
atmosphere, allows one to search within several
astronomical units (AU) of the TW Hydrae stars (an AU is the average
distance between Earth and the sun) for planets
a few times as massive as Jupiter. Newborn planets are quite warm, and
such objects should therefore be sufficiently
luminous to be detected at the distance of this stellar group. In other
words, we should be able to look for newborn
giant planets located at distances from their parent stars similar to
those of giant planets in our own solar system. At
least one brown dwarf, a ``failed star" not massive enough to ignite
hydrogen fusion, has already been found in the TW
Hydrae Association (14), and searches for objects of even lower mass are
under way (15).

The commotion surrounding the TW Hydrae Association has prompted
astronomers to look for other groups like it.
The all-sky survey done by the Roentgen Satellite (ROSAT) has
been particularly useful in identifying isolated
young stars through their x-ray emission. Of the recently discovered
stellar groups, MBM12 and Eta Chamaeleontis
(Eta Cha) are particularly interesting. At about 200 light-years, MBM12 is
the second-nearest group of young stars
after the TW Hydrae Association, containing only 30 to 100 solar masses of
gas. It does not appear to be
gravitationally bound and may be breaking up on a time scale comparable to
the sound-crossing time (16). Thus, in a
few million years, the young stars in MBM12 may appear as isolated objects
not associated with any cloud material,
very similar to how the TW Hydrae stars appear at present. On the basis of
ROSAT detections followed by
ground-based optical spectroscopy, Hearty et al. (17) have identified
eight low-mass young stars associated with
MBM12. Most of them are classical T Tauri stars and are likely to be a
younger population than the TW Hydrae
members. Eta Cha is a cluster of a dozen young stars first identified in
x-ray measurements (18). As with the TW
Hydrae group, Eta Cha is far from any substantial cloud. Its members are
much less dispersed than the TW Hydrae
stars and may represent an epoch intermediate between MBM12 and TW Hydrae
Association.

The exploration of these nearby groups of young stars is progressing at a
breathtaking pace. In the past few months,
telescopes in Arizona, Hawaii, Chile, and Australia were trained on them
with a variety of optical, infrared, and radio
instruments. Many questions remain, but the prospects they offer for
learning about star formation in the solar
neighborhood and the origin and diversity of planetary systems ensure that
interest in them will not wane quickly.

\newpage
{\bf References and Notes}

\noindent 1. G. H. Herbig, in Problems of Physics and Evolution of the Universe, L. V. Mirzoyan, Ed. (Armenian Academy of Science, Yerevan, Armenia, 1978), pp. 171-180\\
\noindent 2. S. M. Rucinski and J. Krautter, Astron. Astrophys. 121, 217 (1983)\\
\noindent 3. R. de la Reza et al., Astrophys. J. 343, L61 (1989); J.
Gregorio-Hetem et al., Astron. J. 103, 549 (1992)\\
\noindent 4. J. H. Kastner et al., Science 277, 67 (1997)\\
\noindent 5. R. A. Webb et al., Astrophys. J. 512, L63 (1999); M. F. Sterzik et al., Astron. Astrophys. 346, L41 (1999)\\ 
\noindent 6. J. R. Stauffer et al., Astrophys. J. 454, 910 (1995); D. R.
Soderblom et al., Astrophys. J. 498, 385 (1998) \\
\noindent 7. E. L. N. Jensen et al., Astron. J. 116, 414 (1998)\\
\noindent 8. E. D. Feigelson, Astrophys. J. 468, 306 (1996) \\
\noindent 9. L. Hartmann, in Star Formation from the Small to the Large Scale, F. Favata, A. A. Kaas, A. Wilson, Eds. (ESA SP-445, European Space Agency, Noordwijk, Netherlands, in press); preprint available at arXiv.org/abs/astro-ph/0001125\\
\noindent 10. S. E. Strom et al., in Protostars and Planets III, E. H. Levy and J.I. Lunine, Eds. (University of Arizona Press, Tucson, AZ, 1993)\\
\noindent 11. R. Jayawardhana et al., Astrophys. J. 521, L129 (1999); R.
Jayawardhana et al., Astrophys. J. 520, L41 (1999); R. D. Gehrz et al., Astrophys. J. 512, L55 (1999)\\
\noindent 12. R. Jayawardhana et al., Astrophys. J. 503, L79 (1998); D.
Koerner et al., Astrophys. J. 503, L83 (1998) \\
\noindent 13. C. M. Telesco et al., Astrophys. J. 530, 329 (2000)\\
\noindent 14. P. J. Lowrance et al., Astrophys. J. 512, L69 (1999) \\
\noindent 15. R. Neuhauser et al., Astron. Astrophys. 354, L9 (2000) \\
\noindent 16. Any dynamical perturbation would travel through the cloud at the
speed of sound. So the sound-crossing time is the typical time scale for 
dynamical evolution of the cloud. \\
\noindent 17. T. Hearty et al., Astron. Astrophys. 353, 1044 (2000)\\ 
\noindent 18. E. E. Mamajek et al., Astrophys. J. 516, L77 (1999)\\

\end{document}